\documentclass[pdflatex,sn-nature]{sn-jnl}

\usepackage{graphicx}%
\usepackage{multirow}%
\usepackage{amsmath,amssymb,amsfonts}%
\usepackage{amsthm}%
\usepackage{mathrsfs}%
\usepackage[title]{appendix}%
\usepackage{xcolor}%
\usepackage{textcomp}%
\usepackage{manyfoot}%
\usepackage{booktabs}%
\usepackage{algorithm}%
\usepackage{algorithmicx}%
\usepackage{algpseudocode}%
\usepackage{listings}%
\theoremstyle{thmstyleone}%
\theoremstyle{thmstyletwo}%

\theoremstyle{thmstylethree}%

\begin{document}

\title{On the coupled geometrical--mechanical origin of the earthquake \textit{b}-value in fault networks}


\author[1,2]{\fnm{Wenbo} \sur{Pan}}

\author*[1]{\fnm{Zixin} \sur{Zhang}}\email{zxzhang@tongji.edu.cn}

\author[2]{\fnm{Björn} \sur{Lund}}

\author*[2]{\fnm{Qinghua} \sur{Lei}}\email{qinghua.lei@geo.uu.se}

\affil[1]{\orgdiv{Department of Geotechnical Engineering, College of Civil Engineering},
\orgname{Tongji University},
\orgaddress{\city{Shanghai}, \country{P.R. China}}}

\affil[2]{\orgdiv{Department of Earth Sciences},
\orgname{Uppsala University},
\orgaddress{\city{Uppsala}, \country{Sweden}}}


\abstract{The Gutenberg--Richter law is a fundamental empirical law in seismology describing earthquake frequency–magnitude distributions, with one of its key parameters, the so-called \textit{b}-value, quantifying the relative frequency of small versus large events. While the \textit{b}-value is commonly interpreted as reflecting crustal heterogeneity and regional stress conditions, its underlying physical origin remains poorly understood, particularly the relative roles of geometrical versus mechanical controls. Here, we develop analytical and numerical models to elucidate the origin of the \textit{b}-value in three-dimensional fault networks subject to mainshock-aftershock sequences. We demonstrate that the \textit{b}-value emerges from the power-law scaling of fault rupture area together with the scaling of slip magnitude. Our results reveal a two-branch frequency–magnitude distribution, with the regime transition governed by fault criticality and fracture energy dissipation, while the transition magnitude reflects the finite population of faults triggered during the sequence. Our findings provide a physically grounded interpretation of earthquake \textit{b}-values, establishing a link between fault mechanics and earthquake statistics.}

\maketitle

\section*{Introduction}\label{sec1}

The Gutenberg--Richter (GR) law describes a fundamental empirical relationship in seismology, whereby the frequency of earthquakes decreases exponentially with increasing magnitude \cite{gutenberg1944frequency}. This relationship is characterized by two key parameters: the \textit{a}-value, reflecting the overall seismic productivity, and the \textit{b}-value, governing the relative proportion of small versus large events. The \textit{b}-value is widely used as a diagnostic indicator of crustal conditions and tectonic settings \cite{de2016statistical,schorlemmer2005variations,scholz2015stress,petruccelli2019influence,herrmann2022revealing}, with reported \textit{b}-values in general ranging from 0.8 to 1.2 for regional tectonic seismicity \cite{frohlich1993teleseismic}. Variability in the \textit{b}-value across different tectonic settings and regions raises a fundamental question about its physical origin and controlling mechanisms.

Two broad schools of thought have been developed to explain the physical origin of the \textit{b}-value. One relates the \textit{b}-value to the geometry of fault systems, which often exhibit fractal patterns and power-law scaling \cite{von1980random,andrews1980stochastic,sornette1991dispersion,king1983accommodation,Scholz1991}. In this view, the \textit{b}-value is linked to the fractal dimension \cite{aki1981probabilistic,hirata1989correlation,chen2006correlation} or the power-law length exponent of fault populations \cite{aki1981probabilistic,turcotte1997fractals,bonnet2001scaling}. The other attributes the \textit{b}-value to mechanical factors, including stress conditions, frictional properties, and rupture dynamics \cite{von1980random,hanks1979b,scholz1968frequency,mcevilly1967earthquake,main1989reinterpretation,goebel2017allows,nanjo2018b,bak1989earthquakes,carlson1989properties}. However, no consensus has been reached yet.

Here, we derive analytical formulations showing that the \textit{b}-value originates from the interplay between the geometric scaling of fault rupture areas and the mechanical scaling of fault slip with rupture size. We further perform direct numerical simulations of mainshock--aftershock sequences in three-dimensional (3D) fault systems, and analyze the resulting aftershock statistics, thereby providing quantitative validation of our analytical framework. We report a two-branch frequency--magnitude distribution and demonstrate that it emerges from the interplay between fault criticality and fracture energy dissipation. 

\section*{Results}\label{sec2}
\subsection*{Analytical formulation}
We analytically derive the \textit{b}-value from the statistical properties of fault patterns and slips. Extensive field observations suggest that fault systems commonly obey a power-law length distribution \cite{turcotte1997fractals,bonnet2001scaling,lei2015new,lei2016tectonic} with the probability density function written as:
\begin{equation}
p\left( l \right) \propto {l^{ - \alpha }},
\tag{1}
\end{equation}
where $l$ is the fault length and $\alpha$ is the power-law exponent. Furthermore, field data have revealed that the maximum fault slip magnitude $\delta_{\rm{max}}$ scales with the fault length according to a power-law relationship \cite{schultz2008dependence,walsh1988analysis}:
\begin{equation}
{\delta _{{\rm{max}}}} \propto {l^\beta },
\tag{2}
\end{equation}
where $\beta$ is the exponent that is related to the rupture mode and energy dissipation characteristics.

We further relate these statistical properties to earthquake magnitudes. The seismic moment $M_0$ is defined as \cite{scholz2019mechanics}:
\begin{equation}
{M_0} = G\bar \delta {A_{\rm{r}}},
\tag{3}
\end{equation}
where $G$ is the shear modulus of host rock, $\bar \delta$ is the slip magnitude averaged over the rupture area $A_{\rm{r}}$, and it is expected \cite{scholz2019mechanics,wesnousky2008displacement} that $\bar \delta \propto \delta_{\rm{max}}$. Assuming faults have a circular shape and are fully ruptured, the rupture area equals to the fault area $A$ such that $A_{\rm{r}}=A\propto l^2$, which yields a moment--length scaling relation $M_0\propto l^{(\beta+2)}$. Applying the conservation of probability under the change of variables \cite{sornette2006critical} from $l$ to $M_0$, the probability density function of seismic moment $M_0$ is derived as:
\begin{equation}
p({M_0}) \propto M_0^{(1 - \alpha )/(\beta  + 2) - 1}.
\tag{4}
\end{equation}
Since the moment magnitude $M_{\rm{w}}$ is related to $M_0$ through \cite{hanks1979moment}:
\begin{equation}
{M_{\rm{w}}} = \frac{2}{3}{\log _{10}}{M_0} - 6.07,
\tag{5}
\end{equation}
we obtain:
\begin{equation}
p\left( {{M_{\rm{w}}}} \right) \propto {10^{\frac{3}{2}{M_{\rm{w}}}{\kern 1pt} (1 - \alpha )/(\beta  + 2)}}.
\tag{6}
\end{equation}
The cumulative frequency--magnitude distribution $N(M_{\rm{w}})=\int_{M_{\rm{w}}}^{+\infty} p(M_{\rm{w}})\,dM_{\rm{w}}$, defined as the number of earthquakes with magnitude greater than or equal to $M_{\rm{w}}$, takes the following form:
\begin{equation}
N\left( {{M_{\rm{w}}}} \right) \propto {10^{\frac{3}{2}{M_{\rm{w}}}(1 - \alpha )/(\beta  + 2)}}.
\tag{7}
\end{equation}
By further comparing with the GR law:
\begin{equation}
N\left( {{M_{\rm{w}}}} \right) \propto {10^{ - b{M_{\rm{w}}}}},
\tag{8}
\end{equation}
we can relate the \textit{b}-value to $\alpha$ and $\beta$ as:
\begin{equation}
b = \frac{3}{2}\left( {\frac{{\alpha  - 1}}{{\beta  + 2}}} \right).
\tag{9}
\label{eq:b_alpha_beta}
\end{equation}

Extensive field observations of two-dimensional outcrop maps of fault traces \cite{bonnet2001scaling,berkowitz1998stereological,piggott1997fractal,bour1998connectivity} reveal that their power-law length exponents typically range from 1.5 to 3.0, with a dominant peak around 2.0. Stereological analysis \cite{bonnet2001scaling,berkowitz1998stereological,darcel2003stereological} further yields that the corresponding exponent $\alpha$ for 3D fault systems lies between 2.5 and 4.0, with a concentration around 3.0. In addition, field data suggest that the exponent $\beta$ commonly falls between 0.5 and 2.0, with many faults displaying nearly linear slip--length scaling (i.e., $\beta\approx 1.0$) \cite{bonnet2001scaling,schultz2008dependence,walsh1988analysis,cowie1992physical}. Substituting the typical values of $\alpha=3.0$ and $\beta=1.0$ into Eq.~(\ref{eq:b_alpha_beta}) yields a \textit{b}-value of 1.0, consistent with commonly reported \textit{b}-values of around 1.0 in earthquake catalogues \cite{taroni2021good,shcherbakov2013aftershock,knopoff1982b,guo1997statistical}. Our analytical solution highlights the fundamental roles of fault network geometry and fault slip behavior, encapsulated by the exponents $\alpha$ and $\beta$, respectively, in governing earthquake statistics.

\subsection*{Numerical simulation}
In natural systems, however, earthquakes develop within a dynamically evolving fault network, where reactivation is governed by multiple interacting processes, such as static and dynamic stress changes \cite{freed2005earthquake}, triggering cascades \cite{palgunadi2024rupture,gabriel2024fault}, and time-dependent frictional state \cite{ikari2013slip,cocco2023fracture}. As a result, faults may not necessarily be fully ruptured, implying conditions that differ from those assumed in the above analytical derivation. To account for these effects and to assess the robustness of our analytical framework for more realistic conditions, we develop numerical models of three-dimensional fault systems undergoing mainshock--aftershock sequences. Although the GR law was originally established from empirical observations of regional seismicity, it has also been shown to hold for aftershock sequences \cite{chen2006correlation,shcherbakov2013aftershock,gulia2019real}, for which reported \textit{b}-values \cite{von1980random,shcherbakov2013aftershock,knopoff1982b,yamashita2003regularity} typically span a broader range of 0.6--1.5.

\begin{figure}
\centering
\includegraphics[width=0.9\textwidth]{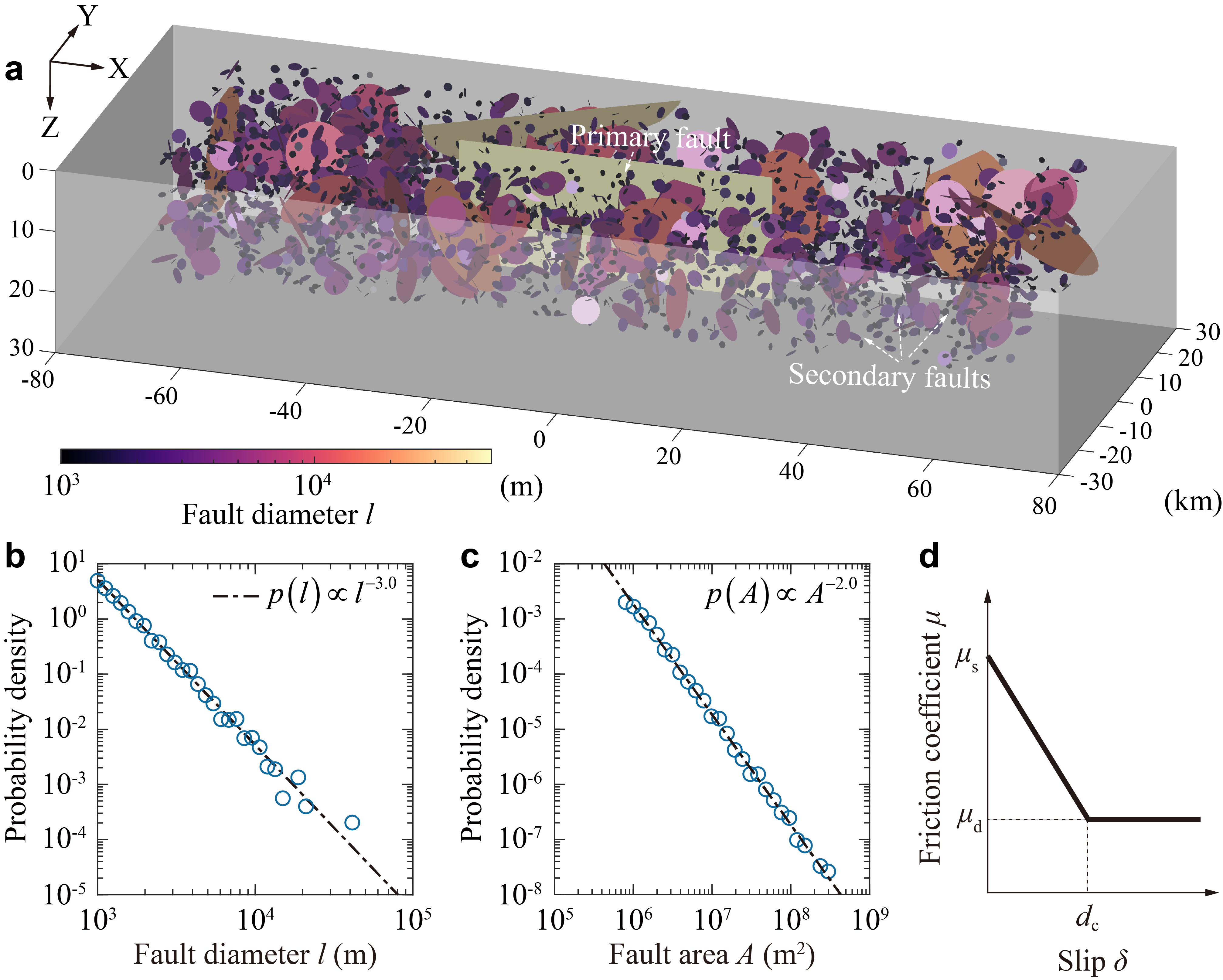}
\caption{\textbf{Numerical model setup and fault network characteristics.} \textbf{a} Model geometry showing a vertical 50 km $\times$ 20 km primary fault surrounded by a network of circular, disk-shaped secondary faults. Probability density functions of \textbf{b} fault diameters and \textbf{c} fault areas, both following power-law scaling. \textbf{d} Fault slip governed by a linear slip-weakening friction law.}
\label{fig1}
\end{figure}

We simulate a mainshock resulting from the dynamic rupture along a 50 km-long strike-slip fault that extends vertically from the ground surface to a depth of 20 km (Fig.~\ref{fig1}a). The primary fault is surrounded by a network of 3,000 randomly oriented secondary faults following a power-law size distribution (Figs.~\ref{fig1}b and \ref{fig1}c), with the exponent $\alpha=3.0$. The secondary faults are disk-shaped, with diameters ranging from 1 to 40 km. The entire fault system is embedded in a crustal rock formation characterized by a Young's modulus $E$ of 70 GPa, a Poisson's ratio $\nu$ of 0.25, and a density $\rho$ of 2,700 kg/m$^{3}$. Both primary and secondary faults are governed by a linear slip-weakening friction law (Fig.~\ref{fig1}d), in which shear failure occurs when the shear-to-normal stress ratio reaches the static friction coefficient ($\mu_{\rm{s}}=0.6$), after which the friction coefficient linearly decreases to the dynamic value ($\mu_{\rm{d}}=0.4$) over a critical slip distance $d_{\rm{c}}$. The primary fault is assigned with $d_{\rm{c}}=1.5$ m, while $d_{\rm{c}}$ for secondary faults is systematically varied from 0.005 m to 5 m across simulations to assess its influence on aftershock behavior. The model is subject to a depth-dependent far-field stress loading (see \nameref{methods}), which produces a maximum shear stress oriented parallel to the strike of the primary fault, thereby promoting left-lateral slip on the primary fault. The shear-to-normal stress ratio on the primary fault is 0.5. Rupture initiates from a nucleation zone centered on the hypocenter due to stress drop and propagates spontaneously along the fault at a sub-Rayleigh speed, reaching approximately 70\% of the shear wave velocity and generating a left-lateral strike-slip mainshock with magnitude $M_{\rm{w}}=7.6$. Secondary faults under coseismic stress changes could be either statically and/or dynamically triggered to rupture once failure conditions are met. Reactivated secondary faults rupture completely only if the shear stress concentration at the advancing rupture front remains above the local frictional strength until the front reaches the fault boundaries; otherwise, rupture arrests when the stress concentration falls below the strength, leaving only part of the fault ruptured.

\begin{figure}
\centering
\includegraphics[width=0.9\textwidth]{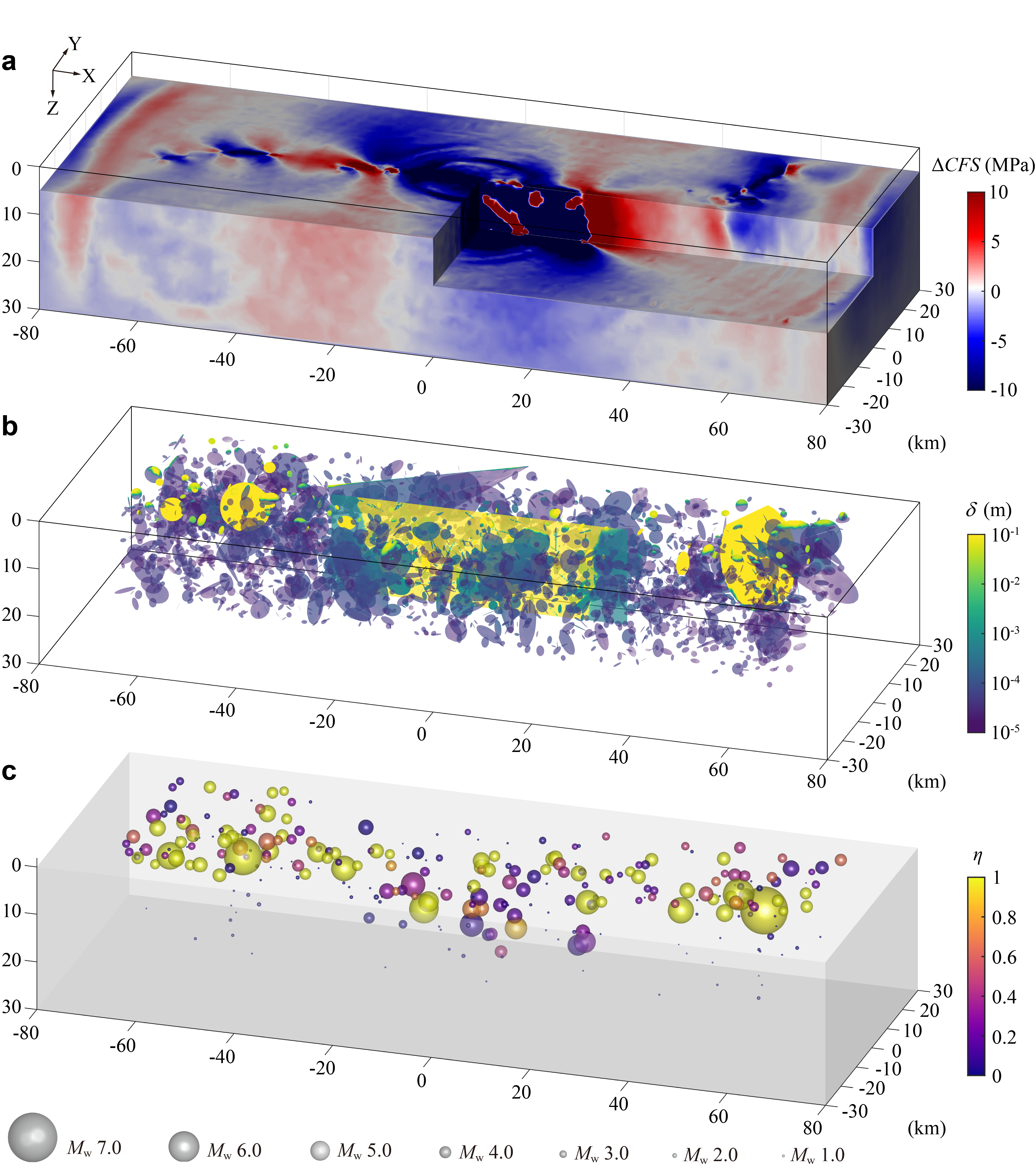}
\caption{\textbf{Coseismic stress changes and aftershock patterns triggered by the mainshock.} \textbf{a} Coulomb failure stress changes ($\Delta$CFS) in the crustal formation induced by the mainshock (through static and dynamic triggering) and further modified locally by aftershock ruptures. \textbf{b} Slip distribution within the fault network. \textbf{c} Spatial distribution of aftershock moment magnitudes, where each marker represents a fault rupture event, with marker size proportional to moment magnitude $M_{\rm{w}}$, marker center indicating the hypocenter, and marker color reflecting the rupture ratio $\eta=A_{\rm{r}}/A$ (where $A$ is total fault area and $A_{\rm{r}}$ is rupture area). All panels show the case with critical slip distance $d_{\rm{c}}=0.5$ m. Aftershocks represent the cumulative events up to 28 s after mainshock initiation, selected as a representative late-time snapshot when the simulated aftershock sequence has converged.}
\label{fig2}
\end{figure}

Coseismic stress perturbations in the crustal formation induced by the mainshock rupture on the vertical primary fault are quantified using Coulomb failure stress changes ($\Delta$CFS) \cite{harris1998introduction}, defined as $\Delta$CFS$=\Delta\tau-\mu_{\rm{s}}\,\Delta\sigma_{\rm{n}}$, where $\Delta\tau$ and $\Delta\sigma_{\rm{n}}$ represent changes in shear and normal stresses, respectively, resolved at each grid point in the numerical model based on the orientation of the primary fault plane (see Fig.~\ref{fig2}a for a representative case with a critical slip distance $d_{\rm{c}}=0.5$ m for secondary faults). Positive $\Delta$CFS values tend to promote aftershock occurrence, whereas negative values are commonly associated with stress shadows and reduced seismicity. In our simulations, the mainshock modifies the surrounding stress field through both static triggering (in which permanent stress redistribution creates localized stress concentrations and relaxations) and dynamic triggering (in which radiated seismic waves induce transient stress fluctuations) (Fig.~\ref{fig2}a; Supplementary Video 1). These coseismic stress changes reactivate portions of the secondary fault network (Fig.~\ref{fig2}b; Supplementary Video 2), giving rise to aftershocks. Fig.~\ref{fig2}c illustrates the spatial pattern of aftershocks for the representative case with $d_{\rm{c}}=0.5$ m. In this scenario, the mainshock has triggered slip on 274 secondary faults, among which 58 undergo complete rupture with three events exceeding magnitude $M_{\rm{w}}\hspace{0.2em}\rm{6.0}$, while others are partially ruptured (Fig.~\ref{fig2}c). The aftershocks show a strong depth-dependent distribution, with more than 60 \% occurring above 5 km and over 75 \% above 10 km.

We project the hypocenters of aftershocks onto a horizontal plane that is superimposed on $\Delta$CFS contours (Supplementary Fig.~1a; Supplementary Video 3). Stress-shadow regions ($\Delta \rm{CFS}<0$) in the direction perpendicular to the primary fault predominantly host small events associated with partially ruptured faults. A limited number of moderate-to-large aftershocks also occur within stress shadows but are closely aligned with the primary fault; these events are likely promoted by strong stress perturbations at intersections between the primary and secondary faults. In contrast, positive $\Delta$CFS lobes near fault tips can trigger fully ruptured faults and generate moderate-to-large events. Notably, moderate-to-large events also occur outside the positive static $\Delta$CFS lobes, where some faults can still rupture fully, due to the triggering by coseismic dynamic waves. Once secondary faults rupture, they further modify the surrounding stress field through dynamically radiated waves and statically changed stress fields (Fig.~\ref{fig2}a; Supplementary Fig.~1a; Supplementary Videos 1 and 3), which may trigger cascading failures or produce localized shadows. We also analyze the case with a smaller critical slip distance, $d_{\rm{c}}=0.005$ m (Supplementary Fig.~1b), where the mainshock triggers 339 aftershocks, including 113 on fully ruptured faults. Moderate-to-large events are preferentially concentrated within the positive $\Delta$CFS lobes near fault tips, whereas stress-shadow regions remain comparatively quiescent. Dynamic stresses can further promote seismicity, as evidenced by more aftershocks occurring within regions of negative static $\Delta$CFS. Notably, some faults that remain inactive or are only weakly activated in the case of $d_{\rm{c}}=0.5$ m become fully ruptured in this case of $d_{\rm{c}}=0.005$ m, with larger events generated. This comparative analysis illuminates the strong control of $d_{\rm{c}}$ on earthquake triggering. A smaller $d_{\rm{c}}$ yields a narrower breakdown zone and stronger stress concentration at the rupture front (Supplementary Fig.~2), thereby facilitating rupture propagation and promoting the development of fully ruptured faults.

\subsection*{Two-branch earthquake frequency–magnitude distribution}

\begin{figure}
\centering
\includegraphics[width=0.9\textwidth]{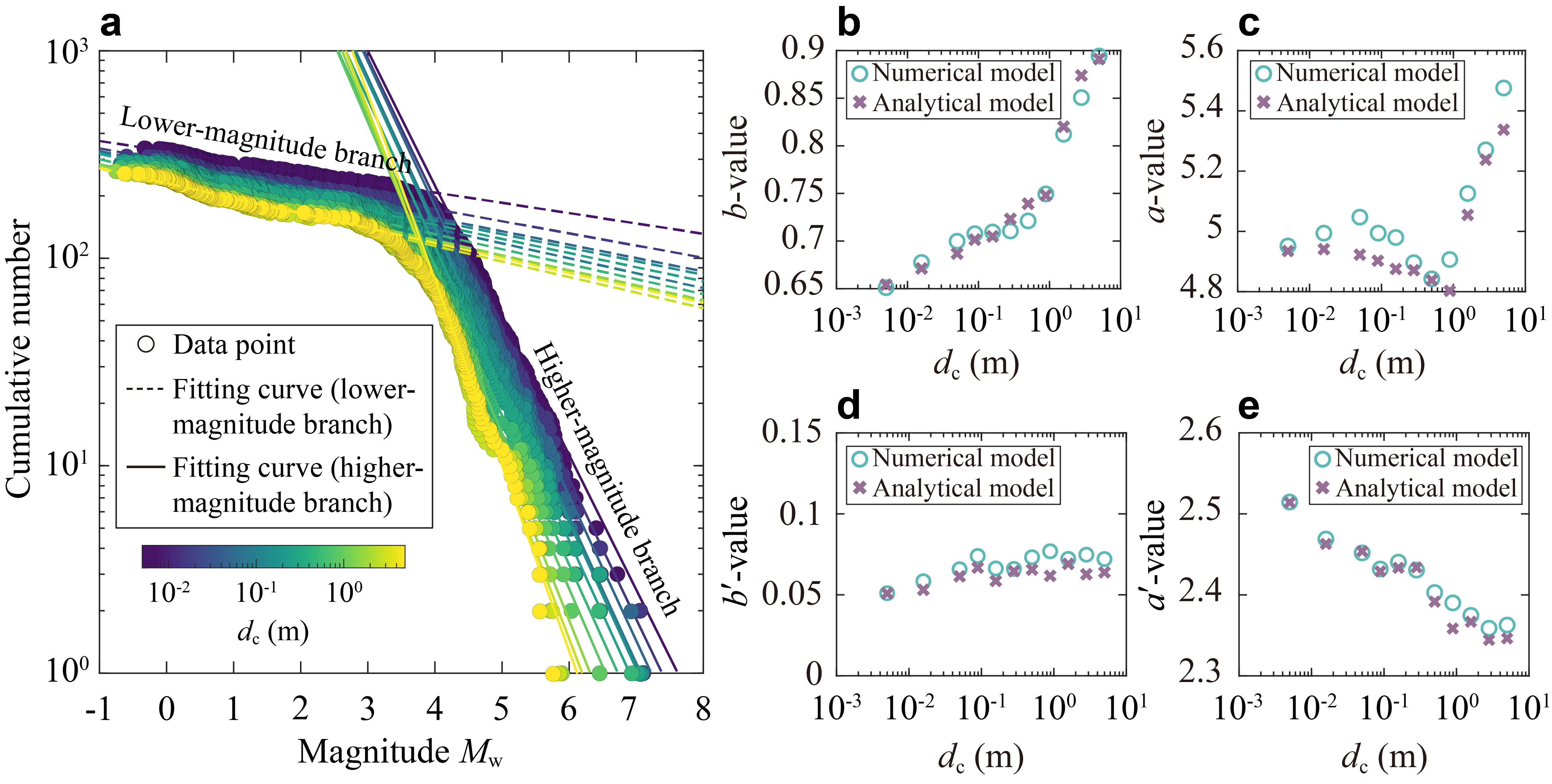}
\caption{\textbf{Two-branch earthquake frequency--magnitude distributions for different values of critical slip distance $d_{\rm{c}}$.} \textbf{a} Cumulative distributions of moment magnitude $M_{\rm{w}}$ for different $d_{\rm{c}}$. Curves and data points are color-coded by the value of $d_{\rm{c}}$, as indicated by the accompanying color map. \textbf{b} \textit{b}-value and \textbf{c} \textit{a}-value of the higher-magnitude branch versus $d_{\rm{c}}$. \textbf{d} $b^{\prime}$-value and \textbf{e} $a^{\prime}$-value of the lower-magnitude branch versus $d_{\rm{c}}$. In \textbf{b--e}, numerical estimates are compared with analytical predictions obtained from the measured scaling exponents: \textit{b} and \textit{a} derived from $\alpha$ and $\beta$, and $b^{\prime}$ and $a^{\prime}$ derived from $\alpha^{\prime}$ and $\beta^{\prime}$.}
\label{fig3}
\end{figure}

The cumulative frequency--magnitude distributions of simulated aftershocks for varying $d_{\rm{c}}$ are shown in Fig.~\ref{fig3}a. These distributions generally conform to the GR law but reveal a distinct scaling break at moment magnitudes $M_{\rm{T}} \approx 4.0$--4.5, thereby exhibiting a two-branch behavior. Here, $M_{\rm{T}}$ denotes the transition magnitude over which aftershock magnitudes follow the higher-magnitude branch. The lower-magnitude branch has a gentler slope, denoted by the $b^{\prime}$-value, whereas the higher-magnitude branch follows a steeper slope corresponding to the conventional \textit{b}-value. The two branches are also characterized by distinct intercepts, denoted by $a^{\prime}$ and \textit{a}, respectively.

We estimate the parameters \textit{b}, $b^{\prime}$, \textit{a}, and $a^{\prime}$ using the maximum likelihood estimation method, with results shown in Figs.~\ref{fig3}b--e. We first analyze the trends in the numerical simulation results. As $d_{\rm{c}}$ increases, the \textit{b}-value exhibits a gradual increase at small $d_{\rm{c}}$ followed by a more pronounced rise once a threshold of around 1 m is exceeded. This reflects a shift in rupture dynamics: beyond this threshold, rupture and slip on all faults in the system become increasingly constrained due to insufficient driving energy, restricting their potential to propagate and grow (see Supplementary Note 1). A similar behavior is observed for the corresponding \textit{a}-value, which initially shows relatively small fluctuations but grows substantially beyond such a threshold. For the lower-magnitude branch, the $b^{\prime}$-value increases slightly before plateauing, whereas the $a^{\prime}$-value continuously declines with increasing $d_{\rm{c}}$, indicating a progressive reduction in the total number of aftershocks.

The numerical results further show that rupture area scales with cumulative frequency and that slip scales with rupture area following two-branch power-law relations, from which we extract the exponents $\alpha^{\prime}$ and $\beta^{\prime}$ for the lower-magnitude branch, and $\alpha$ and $\beta$ for the higher-magnitude branch (Supplementary Note 2 and Supplementary Figs.~3--5). Although the fault network geometry is fixed (including the prescribed power-law length exponent of 3.0), the rupture statistics are not. Varying $d_{\rm{c}}$ alters rupture arrest and propagation, thereby modifying the emergent distributions of rupture area and slip, and yielding distinct values of $\alpha^{\prime}$, $\beta^{\prime}$, $\alpha$, and $\beta$. Substituting these measured exponents into our analytical formulation (Eq.~(\ref{eq:b_alpha_beta}) and Eq.~(S4)) yields $a^{\prime}$, $b^{\prime}$, \textit{a}, and \textit{b} values that closely match those obtained directly from the frequency--magnitude fits (Figs.~\ref{fig3}b--e). The robustness of our results is confirmed by additional simulations in fault networks with different power-law length exponents (Supplementary Figs.~6 and 7). Collectively, these results indicate that the \textit{b}-value is controlled by the geometrical scaling of fault rupture area and the mechanical scaling of slip with rupture area.

\begin{figure}
\centering
\includegraphics[width=0.9\textwidth]{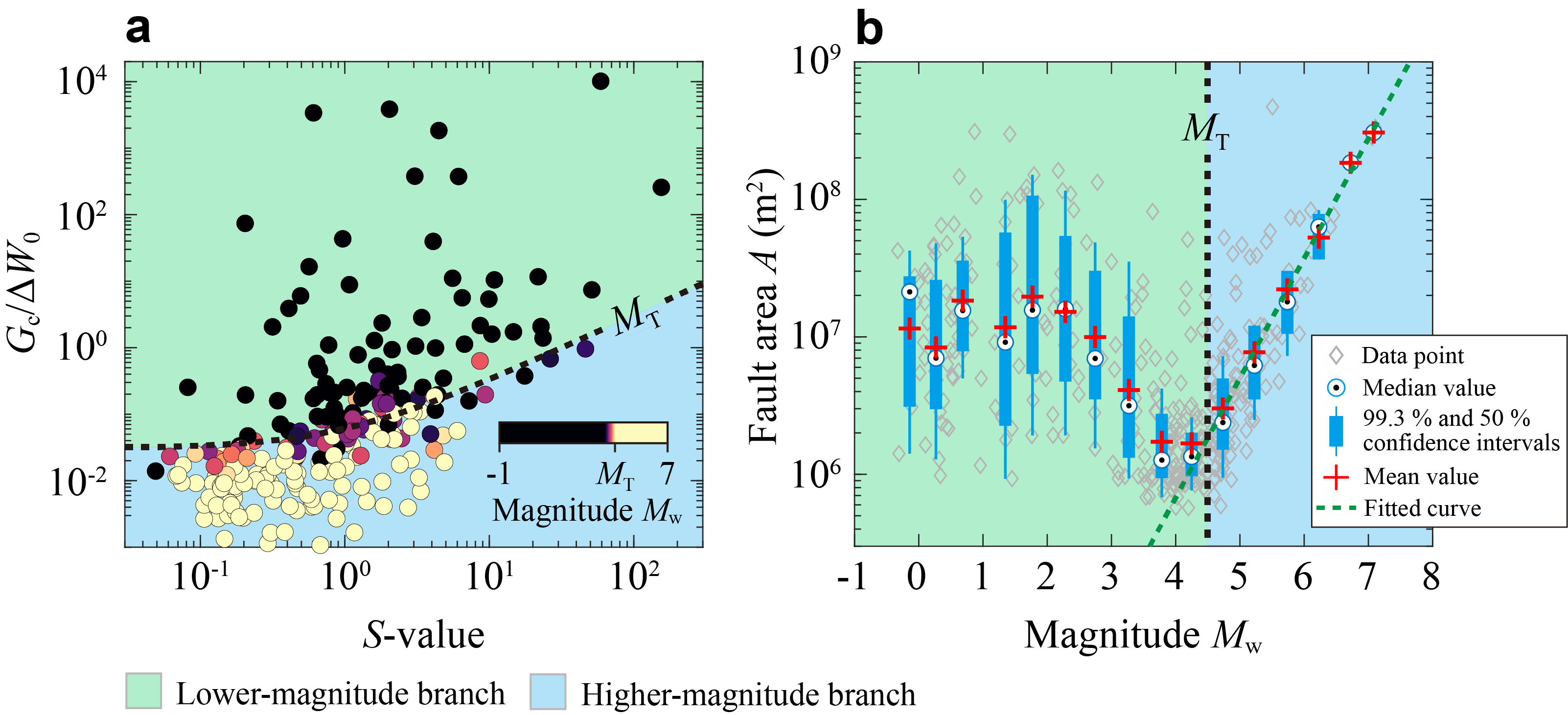}
\caption{\textbf{Transitions in rupture regimes in the fault network.} \textbf{a} Energy ratio $G_{\rm{c}}/\Delta W_{0}$, which quantifies the proportion of available energy dissipated as fracture energy, plotted against the \textit{S}-value that characterizes fault stress criticality. Symbols show simulation results for critical slip distance $d_{\rm{c}}=0.005$ m. The dashed curve marks the regime boundary given by the stress--energy formulation, Eq.~(\ref{eq:energy_framework_with_M_0}). Here, we identify the transition magnitude $M_{\rm{T}}$ from the cumulative frequency--magnitude distribution of this simulation case ($M_{\rm{T}}=4.5$), convert it to the corresponding seismic moment ($M_{0}=7.16\times 10^{15}$ N$\cdot$m), and substitute this $M_{0}$ into Eq.~(\ref{eq:energy_framework_with_M_0}) to obtain the delineation. \textbf{b} Relationship between fault area $A$ and moment magnitude $M_{\rm{w}}$ for $d_{\rm{c}}=0.005$ m, revealing a transition from fault area-independent behavior in the lower-magnitude branch to a fault area-dependent behavior in the higher-magnitude branch.}
\label{fig4}
\end{figure}

The two-branch scaling observed in the cumulative frequency--magnitude distribution can be mechanistically explained by the interplay of two fundamental controlling parameters (Fig.~\ref{fig4}a). The first is the \textit{S}-value \cite{andrews1976rupture,das1977numerical,liu2014progression}, defined as $S=(\bar{\tau}_{\rm{s}}-\bar{\tau}_{0})/(\bar{\tau}_{0}-\bar{\tau}_{\rm{f}})$, where $\bar{\tau}_{\rm{s}}$, $\bar{\tau}_{0}$, and $\bar{\tau}_{\rm{f}}$ respectively denote the static, initial, and final shear stresses averaged over the rupture area. The \textit{S}-value quantifies fault criticality, characterizing how close the initial shear stress $\bar{\tau}_{0}$ is with respect to the static shear strength $\bar{\tau}_{\rm{s}}$, normalized by the static stress drop $\bar{\tau}_{0}-\bar{\tau}_{\rm{f}}$. The second parameter is the energy ratio $G_{\rm{c}}/\Delta W_{0}$, which quantifies the fraction of the available energy $\Delta W_{0}$ dissipated as fracture energy $G_{\rm{c}}$ during rupture propagation. The available energy $\Delta W_{0}$ represents the portion of the total released elastic strain energy $W_{\rm{F}}$ that is not dissipated as frictional heat $E_{\rm{H}}$ and is therefore available to drive rupture growth \cite{scholz2019mechanics,cocco2023fracture,kaneko2017slip,kammer2024earthquake}. It can be expressed as $\Delta W_{0}=W_{\rm{F}}-E_{\rm{H}}=1/2 (\bar{\tau}_{0}+\bar{\tau}_{\rm{f}}-2\bar{\tau}_{\rm{d}})\,\bar{\delta}A_{\rm{r}}$ (Supplementary Fig.~8), where $\bar{\tau}_{\rm{d}}$ is the dynamic shear stress averaged over the rupture area. The fracture energy itself is given by $G_{\rm{c}}=1/2 (\bar{\tau}_{\rm{s}}-\bar{\tau}_{\rm{d}})\,d_{\rm{c}}\,A_{\rm{r}}$. Notably, the \textit{S}-value and the energy ratio $G_{\rm{c}}/\Delta W_{0}$ are related. Under an assumption of $\bar{\tau}_{\rm{d}}=\bar{\tau}_{\rm{f}}$, one can obtain:
\begin{equation}
\frac{{{G_{\rm{c}}}}}{{\Delta {W_{\rm{0}}}}} = \left( {S + 1} \right)\frac{{{d_{\rm{c}}}}}{{\bar \delta }},
\tag{10}
\label{eq:energy_framework}
\end{equation}
indicating that more critically stressed faults (lower \textit{S}-values) dissipate a smaller fraction of the available energy as fracture energy, thereby favoring more extensive rupture propagation. Substituting the equation $M_0=G\bar{\delta}A_{\rm{r}}$ along with the relation between $\bar{\delta}$ and $A_{\rm{r}}$ (see Eq.~(S3) in Supplementary Note 2) into Eq.~(\ref{eq:energy_framework}) yields:
\begin{equation}
\frac{{{G_{\rm{c}}}}}{{\Delta {W_{\rm{0}}}}} = \left( {S + 1} \right){d_{\rm{c}}}{\left( {\frac{G}{{{M_0}K_2^{2/\beta }}}} \right)^{\beta /\left( {\beta  + 2} \right)}},
\tag{11}
\label{eq:energy_framework_with_M_0}
\end{equation}
where $K_2$ is the scaling coefficient in the relation between $\bar{\delta}$ and $A_{\rm{r}}$ (see Supplementary Note 2).

We then construct a phase diagram using the \textit{S}-value and energy ratio $G_{\rm{c}}/\Delta W_{0}$ (Fig.~\ref{fig4}a), where two distinct rupture regimes are identified. The regime boundary (dashed curve) corresponds to the transition magnitude $M_{\rm{T}}$ that marks the break in the frequency--magnitude scaling. For the higher-magnitude branch, faults are more critically stressed (low \textit{S}-values) and require less energy to initiate and sustain rupture propagation (low $G_{\rm{c}}/\Delta W_{0}$). Under such conditions, rupture propagates spontaneously and self-sustains, enabling large, cascading events. In this regime, aftershock magnitude $M_{\rm{w}}$ scales systematically with fault area $A$, reflecting strong geometrical control exerted by the fault network (Fig.~\ref{fig4}b). Conversely, faults in the lower-magnitude branch are generally less critically stressed (high \textit{S}-value), and a substantial portion of the available energy is consumed in overcoming fracture energy barriers to propagate rupture (high $G_{\rm{c}}/\Delta W_{0}$). In many cases, the energy budget is insufficient to sustain rupture growth, leading to premature arrest and limited rupture extent. As a consequence, ruptures in this regime are not geometrically constrained, and aftershock magnitudes show little dependence on fault area (Fig.~\ref{fig4}b). This transition between rupture regimes gives rise to the break in the frequency--magnitude distribution, marking a crossover between two distinct scaling branches.

\section*{Discussion}
Prior studies \cite{king1983accommodation,aki1981probabilistic,chen2006correlation,legrand2002fractal} have proposed a geometrical interpretation of the \textit{b}-value, linking it to the fractal dimension $D$ through the relation $D=3b/c$, where $c$ is the slope of $\log_{10} M_0$ versus $M_{\rm{w}}$ and is typically close to 1.5.\cite{kanamori1978quantification} This relationship can be recovered as a special case of our more general analytical expression, $b=\frac{3}{2}\left(\frac{\alpha-1}{\beta+2}\right)$, by assuming self-similar fault systems with $D=\alpha-1$,\cite{king1983accommodation,bonnet2001scaling,lei2016tectonic,turcotte1986fractal,bour2002statistical,davy2010likely} and a linear slip behavior with $\beta=1.0$.\cite{bonnet2001scaling,schultz2008dependence,cowie1992physical} Notably, seismological observations indicate $c\approx 1.0$ for smaller events \cite{chen2006correlation,legrand2002fractal}, implying $\beta\approx 2.5$, consistent with the range of 2.0--2.7 obtained from our simulations for the lower-magnitude branch (see Supplementary Fig.~5b). Hence, our analytical model provides a unified formulation for the \textit{b}-value, incorporating both geometrical and mechanical effects. This formulation naturally accounts for datasets that depart from a strictly positive $D$--\textit{b} correlation, including reported negative correlations \cite{hirata1989correlation,henderson1992evolution}. According to our framework, a relatively small fractal dimension $D$ together with a sufficiently small $\beta$ can still produce a large \textit{b}-value, explaining why high \textit{b}-values were observed in fault networks with small $D$.\cite{hirata1989correlation,henderson1992evolution}

In parallel, other studies have derived the \textit{b}-value from mechanical principles using Burridge-Knopoff-type models with velocity-weakening friction \cite{bak1989earthquakes,carlson1989properties,burridge1967model,carlson1991intrinsic,mori2006simulation,kawamura2017statistical}. In such spring-block systems, stick-slip instabilities can self-organize into a critical state, producing scale-invariant event-size distributions consistent with the GR law \cite{bak1989earthquakes,carlson1989properties}. However, when frictional weakening is pronounced, these systems rupture more coherently, giving rise to quasi-periodic large events rather than a power-law distribution \cite{carlson1989properties,mori2006simulation,kawamura2017statistical}. This limitation likely arises from the lack of fractal heterogeneity in idealized block-spring models, which otherwise suppress coherent rupture. By integrating geometric power-law scaling with frictional instability, we show that the GR-like behavior persists across fault systems with both strong and weak frictional instability.

Taken together, our results indicate that the \textit{b}-value reflects the combined influence of fault geometry and rupture mechanics, rather than either control alone. We further suggest that the two-branch frequency--magnitude scaling behavior may arise from the interplay between fault criticality and rupture dynamics. We also note that the transition magnitude $M_{\rm{T}}$, which delineates the two branches, scales with a characteristic fault area $A_{\rm{min}}$ (Supplementary Note 3; Supplementary Figs.~9 and 10). This area may represent the lower bound associated with the smallest faults within the finite reactivated population that can be triggered by the mainshock \cite{palgunadi2024rupture,gabriel2024fault}.

\section*{Methods}\label{methods}
\subsection*{Numerical modeling framework}

Dynamic rupture simulations are performed using the 3D distinct element code (3DEC) \cite{itasca3dec2020} that explicitly resolves fault interactions, rupture propagation, and stress transfer in faulted rock masses, and has been extensively validated and applied to modeling seismic and coseismic processes \cite{falth2015simulating,cappa2022fluid}. The model domain (Fig.~\ref{fig1}a) is discretized into deformable polyhedral blocks, each further subdivided into tetrahedral constant-strain finite-difference elements to resolve the internal stress--strain evolution. Faults are explicitly represented as assemblies of contact interfaces between adjacent blocks, which are discretized into multiple sub-contacts to resolve both normal and shear deformation along fault surfaces; these sub-contacts coincide with the faces of the finite-difference elements. At the block scale, relative motion is resolved using the discrete element method, with block translation and rotation governed by Newton's second law. The translational motion of a block centroid is described by:
\begin{equation}
m\ddot{\mathbf{x}} = \mathbf{F} - m\zeta \dot{\mathbf{x}},
\tag{12}
\end{equation}
where $\mathbf{x}$ is the centroid displacement vector, $m$ is the block mass, $\zeta$ is a mass-proportional damping coefficient, and $\mathbf{F}$ denotes the resultant force acting on the block, including contact interactions, externally applied loads, and gravitational body forces. Rotational motion follows:
\begin{equation}
I\dot{\pmb{\omega}} = \mathbf{M} - I\zeta \pmb{\omega},
\tag{13}
\end{equation}
where $\pmb{\omega}$ is the angular velocity vector, $\mathbf{M}$ is the resultant torque, and $I$ is the effective moment of inertia. Within each block, internal deformation is resolved using the finite difference method \cite{itasca3dec2020,cundall1988formulation}. The motion of gridpoints (vertices of tetrahedral elements) obeys:
\begin{equation}
m\ddot{\mathbf{u}} = \int_{\partial \Omega} \pmb{\sigma}\cdot \mathbf{n}\,ds + \mathbf{F},
\tag{14}
\end{equation}
where $\mathbf{u}$ is the gridpoint displacement vector, $\pmb{\sigma}$ is the Cauchy stress tensor, $\mathbf{n}$ is the outward unit normal to the boundary $\partial \Omega$ of the control volume $\Omega$, and the surface integral represents the resultant internal force arising from stress tractions within adjacent tetrahedral elements. The force term $\mathbf{F}$ includes external loads, sub-contact forces acting on boundary gridpoints, and gravitational contributions. Strain increments within each element are computed from gridpoint velocity gradients as:
\begin{equation}
\Delta \pmb{\varepsilon} = \frac{\Delta t}{2}\left[ \nabla \dot{\mathbf{u}} + \left(\nabla \dot{\mathbf{u}}\right)^{\rm T} \right],
\tag{15}
\end{equation}
where $\Delta t$ denotes the time step in the numerical simulation. Stress increments are evaluated using a linear isotropic elastic constitutive relation:
\begin{equation}
\Delta \pmb{\sigma} = \lambda \Delta \varepsilon_{v}\mathbf{I} + 2G\Delta \pmb{\varepsilon},
\tag{16}
\end{equation}
where $\lambda$ and $G$ are respectively the first and second Lam\'e constants, $\varepsilon_{v}=\mathrm{tr}(\pmb{\varepsilon})$ is the volumetric strain, and $\mathbf{I}$ is the second-order identity tensor. Normal displacement along fault interfaces (Supplementary Fig.~11a) is described by a linear stress--displacement relation with a tensile cutoff:
\begin{equation}
\sigma_{\rm{n}}=
\begin{cases}
0, & u_{\rm{n}} < 0,\\
k_{\rm{n}}u_{\rm{n}}, & u_{\rm{n}} \ge 0,
\end{cases}
\tag{17}
\end{equation}
where $\sigma_{\rm{n}}$ is the normal stress, $k_{\rm{n}}$ is the normal stiffness, and $u_{\rm{n}}$ is the normal displacement (positive for closure). Shear displacement (Supplementary Fig.~11b) follows a Coulomb-type relation:
\begin{equation}
\tau=
\begin{cases}
0, & u_{\rm{n}} < 0,\\
k_{\rm{s}}u_{\rm{s}}, & u_{\rm{n}} \ge 0, \; u_{\rm{s}} < u_{\rm{p}},\\
\mu \sigma_{\rm{n}}, & u_{\rm{n}} \ge 0, \; u_{\rm{s}} \ge u_{\rm{p}},
\end{cases}
\tag{18}
\end{equation}
where $\tau$ is the shear stress, $k_{\rm{s}}$ is the shear stiffness, $u_{\rm{s}}$ is the shear displacement, and $u_{\rm{p}}$ is the peak elastic displacement. The friction coefficient $\mu$ follows a linear slip-weakening law:
\begin{equation}
\mu(\delta)=
\begin{cases}
\mu_{\rm{s}}-\left(\mu_{\rm{s}}-\mu_{\rm{d}}\right)\delta/d_{\rm{c}}, & \delta < d_{\rm{c}},\\
\mu_{\rm{d}}, & \delta \ge d_{\rm{c}},
\end{cases}
\tag{19}
\label{eq:slip_weakening}
\end{equation}
where $\delta=u_{\rm{s}}-u_{\rm{p}}$ is the slip magnitude, $\mu_{\rm{s}}$ and $\mu_{\rm{d}}$ are the static and dynamic friction coefficients, respectively, and $d_{\rm{c}}$ is the critical slip distance. An explicit time-marching scheme \cite{itasca3dec2020} is adopted to ensure consistent coupling between contact forces and internal stresses computed using the discrete element and finite difference methods, respectively. The model employs mesh sizes ranging from 200 m (minimum) to 500 m (maximum), balancing numerical accuracy and computational efficiency. Apart from the free upper surface representing the ground surface, non-reflecting boundary conditions are applied to all sides of the model to absorb outgoing seismic waves and minimize artificial reflections \cite{falth2015simulating}, approximating an unbounded medium.

\subsection*{Simulation workflow and dynamic rupture calculation}
The simulations proceed in two stages: (i) establishing an initial stress equilibrium and (ii) modeling the mainshock rupture and subsequent aftershock triggering. In the first stage, a far-field tectonic stress state is imposed at model boundaries and a static calculation is carried out to attain mechanical equilibrium. A left-lateral strike-slip event along the primary fault is simulated by imposing the following symmetric stress tensor:
\begin{equation}
\left[
\begin{array}{ccc}
\sigma_{xx} & \tau_{xy} & \tau_{xz} \\
            & \sigma_{yy} & \tau_{yz} \\
\mathrm{sym}. &           & \sigma_{zz}
\end{array}
\right]
=
\left[
\begin{array}{ccc}
\rho gz & 0.5\rho gz & 0 \\
        & \rho gz    & 0 \\
\mathrm{sym}. &       & \rho gz
\end{array}
\right],
\tag{20}
\end{equation}
where $\sigma_{ii}$ ($i=x,y,z$) are the normal stresses, $\tau_{ij}$ ($i\ne j$) are the shear stresses, $\rho$ is the rock density, $g$ is the gravitational acceleration, and $z$ is the depth below the ground surface. Under the far-field stress loading, faults with different orientations are subjected to heterogeneous shear-to-normal stress ratios, and thus exhibit variable stress criticality (\textit{S}-values). Interactions among neighboring faults further enhance this variability within the fault system. In the second stage, dynamic rupture nucleation and propagation are governed by the linear slip-weakening friction law (Eq.~(\ref{eq:slip_weakening})). Rupture is initiated by imposing a localized 0.5 MPa stress drop at a predefined hypocenter within a 3 km-radius critically-stressed nucleation zone \cite{ohnaka2000physical} (Supplementary Fig.~12), where the shear stress is set equal to the shear strength by assigning a static friction coefficient of 0.5, matching the primary fault's initial shear-to-normal stress ratio. The rupture then propagates spontaneously along the primary fault plane and arrests at the fault boundaries. To avoid spurious rupture arrest near the primary fault tips, both $\mu_{\rm{s}}$ and $\mu_{\rm{d}}$ are increased linearly to 1.0 within the outermost 5 km of the primary fault, reducing stress concentrations and ensuring a physically consistent rupture termination associated with the mainshock. Secondary faults are reactivated when their local shear-to-normal stress ratio reaches the failure criterion ($\tau/\sigma_{\rm{n}}=\mu_{\rm{s}}$), either directly from mainshock-induced stress changes or indirectly through stress transfer from neighboring faults. These secondary fault ruptures propagate spontaneously under slip-weakening friction, and arrest either at fault boundaries (corresponding to the fully ruptured scenario) or when the driving force is insufficient to overcome the resisting force (corresponding to the partially ruptured scenario).

\bibliography{references_V2}

\section*{Acknowledgements}
Z.Z. acknowledges support from the National Natural Science Foundation of China (grant no.~42377146). Q.L. is grateful for support from the European Research Council (ERC) under the European Union’s Horizon Europe programme (ERC Consolidator Grant, grant no.~101232311) for the project “Unified framework for modelling progressive to catastrophic failure in fractured media (FORECAST)”.

\section*{Author contributions}
W.P. contributed to conceptualization, methodology, software, formal analysis, investigation, and drafted the manuscript. Z.Z., B.L., and Q.L. contributed to conceptualization, investigation, resources, supervision, project administration, and critically revised the manuscript. Q.L. contributed to methodology and co-wrote the original draft. Z.Z. and Q.L. acquired funding.

\section*{Competing interests}
The authors declare no competing interests.

\end{document}